\documentclass[pre,groupedaddress,twocolumn,showpacs,amsmath,floatfix,a4paper]{revtex4}

\usepackage{graphicx}

\newcommand{\nn}     {\hat{\mathbf{n}}}
\newcommand{\rr}     {\mathbf{r}}
\newcommand{\ff}     {\mathbf{f}}
\newcommand{\ee}     {\mathrm{e}}

\newcommand{\uu}     {\mathbf{u}}
\newcommand{\vv}     {\mathbf{v}}
\newcommand{\RR}     {\mathbf{R}}
\newcommand{\VV}     {\mathbf{V}}
\newcommand{\PP}     {\mathbf{P}}
\newcommand{\FF}     {\mathbf{F}}

\begin{document}
\title{Efficient simulation of non-crossing fibers and chains in a hydrodynamic solvent}
\author{J.~T.~Padding}

\affiliation{
  Computational~Biophysics,
  Faculty of Science and Technology, University~of~Twente,
  P.O.~Box~217, 7500~AE~Enschede, The~Netherlands}
\date{\today}

\begin{abstract}
An efficient simulation method is presented for Brownian fiber 
suspensions, which includes both uncrossability of the fibers and hydrodynamic
interactions between the fibers mediated by a mesoscopic solvent. To conserve hydrodynamics, collisions between the fibers are treated such
that momentum and energy are conserved locally. The choice of simulation parameters is rationalised on the basis of dimensionless numbers
expressing the relative strength of different physical processes.
The method is applied to suspensions of semiflexible fibers with a contour length equal to the persistence length,
and a mesh size to contour length ratio ranging from 0.055 to 0.32. 
For such fibers the effects of hydrodynamic interactions are observable, but relatively small. The non-crossing constraint, on the other hand,
is very important and leads to hindered displacements of the fibers, with an effective tube diameter in agreement with recent
theoretical predictions. The simulation technique opens the way to study the effect of viscous effects and hydrodynamic interactions in
microrheology experiments where the response of an actively driven probe bead in a fiber suspension is measured.

\end{abstract}


\maketitle

\section{\label{intro}Introduction}

The dynamics of rods and semiflexible fibers are strongly influenced by their mutual uncrossability.
Examples include carbon nanotubes \cite{Rafii04}, fd-virus \cite{Lettinga04,Kang06,Holmqvist07,Kang07},
and biologically relevant polymers such as actin \cite{Kas94,Kas96,Brangwynne07,Semmrich08}
and tubulin \cite{Jones05,Lin07}.
Already at surprisingly low concentrations, uncrossability in such systems leads to a temporary and anisotropic ``cage''
or tube from which the rod or fiber can only escape through anisotropic motion (reptation) \cite{DoiEdwards}
or through collective motion, as exemplified by the collective reorientation observed in sheared concentrated
rod suspensions \cite{Tao05,Tao06,Ripoll08}.

Besides the mutual uncrossability constraint, the dynamics of rods and fibers are also influenced by Brownian forces
(due to random collisions with solvent molecules) and hydrodynamic interactions (HIs) mediated by the solvent.
The role of HIs in entangled suspensions of Brownian rigid rods and semiflexible fibers has remained, with a few exceptions,
largely unexplored. This is caused by the difficulty of treating Brownian dynamics, hydrodynamics, and entanglements within one theoretical
framework \cite{DoiEdwards,Dhont}.
HIs are dominant in the dilute and onset of the semidilute regime. For example, the scaling of the relaxation times of the normal modes
(Rouse modes) in an unentangled bead-spring chain changes from $\tau_p \propto (N/p)^2$ for a chain without HIs to
$\tau_p \propto (N/p)^{3/2}$ for a chain with HIs (Zimm scaling) \cite{DoiEdwards,Winkler04}. Here $N$ is the number
of beads and mode $p$ measures correlated motion on a length scale of $(N/p)$ beads.
Also the diffusion and segmental dynamics of dilute DNA molecules are controlled by hydrodynamic interactions \cite{Winkler06,Petrov06}.
On the other hand,
it is believed that HIs are effectively screened in very concentrated
suspensions and to a certain extent also in semidilute suspensions
in equilibrium situations \cite{DoiEdwards,Pryamitsyn08,Kang07}.
The onset of the semidilute regime already occurs at lower concentrations for rigid rods than for flexible chains of equal contour
length \cite{DoiEdwards}. This corresponds to a smaller dynamic correlation length in a semidilute suspension of rigid rods than in an equally
concentrated suspension of flexible chains. Indeed,
Pryamitsyn and Ganesan have shown that the effects of HIs in semidilute and concentrated suspensions of
completely rigid Brownian rods (with aspect ratio up to 20) are secondary relative to the steric interactions \cite{Pryamitsyn08}.
A detailed analysis shows that
HIs modify the diffusion parallel to the rod, in agreement with theories of hydrodynamic screening \cite{Muthukumar83,Shaqfeh90}.
In all probability, the importance of HIs is decreasing with increasing chain stiffness and/or increasing concentration,
but it is difficult to predict in general under which conditions HIs can be neglected.

The need to consider HIs becomes particularly important when considering non-equilibrium situations. There are various applications where fibers
are dragged along by flow or where the fibers generate flow because they are dragged by an external field. Examples include flow through
microchannels \cite{Shaqfeh06}, sedimentation or electrophoresis of fibers \cite{Butler02,Mackaplow98,Llopis07},
and active microrheology \cite{Valberg87,Zaner89,Wiggins98,Bausch99}. In active microrheology a colloidal bead is embedded in a medium and driven
by magnetic or optical forces. The force-displacement response is measured with the goal to
locally measure the rheology of the medium. In case of a medium consisting of a fiber network, it is important for the interpretation of
these experiments to understand the hydrodynamic coupling between fluid flow generated by the probe bead on the one hand, and the fiber network
on the other hand.
The work presented here is part of a long-term effort to generate this understanding.
Coupling between fluid flow and fiber dynamics may be especially important when the probe bead is smaller than the mesh size of the network \cite{Kang07}.
Even for probe sizes in between the mesh size and the fiber contour length interesting new mechanisms may be observed \cite{Pryamitsyn08b}.

Computer simulations in which HIs, entanglements, and Brownian motion are treated on an equal footing may help in gaining insight in the dynamics of
Brownian fiber suspensions.
First, let us focus on hydrodynamics.
To rigorously include HIs in a simulation requires a decomposition of the mobility tensor, which is typically an $\mathcal{O}(N^3)$
operation \cite{Ermak78}, although with certain approximations (expanding force distributions along rods in Legendre polynomials and
retaining only lower order terms) this can be made more efficient \cite{Shaqfeh06,Butler02,Butler05}.
Another approach is to explicitly include the solvent. The large gap in time- and lengthscales between the solvent molecules and colloidal sized particles
has led to the development of mesoscopic simulation techniques which avoid the computationally costly explicit treatment of every solvent molecule.
Important developments in this area are Lattice Boltzmann (when extensions to allow for
thermal fluctuations are included) \cite{Succi,Ladd01,Cates04,Usta05},
Dissipative Particle Dynamics (DPD) \cite{Hoogerbrugge92,Espanol95}, and Multi-Particle Collision Dynamics (MPCD)
\cite{Malevanets99,Ihle01,IhleKroll03,Kikuchi03,Winkler04,Padding04,Padding05,Ihle05,Pooley05,Ripoll05,Padding06,Goetze07,Watari07,Padding08}.
The latter, in its original implementation \cite{Malevanets99}, is also known as Stochastic Rotation Dynamics (SRD).
All these mesoscopic simulation techniques account for correlated motion of the solvent which leads to long-range hydrodynamic interactions.

Second, let us focus on the entanglements. Most existing methods implement non-crossing by resorting to explicit repulsive interactions.
The dynamics of relatively short non-crossing rods may be modeled by means of forcefields with ellipsoidal or spherocylindrical
geometry \cite{Demiguel91,Jose04}, whereas non-crossing rods or chains are often modeled by representing them as a string of relatively hard beads with
bonds that are sufficiently strong to make the crossing of two such chains energetically unfavourable \cite{Kremer90,Kroger04,Pryamitsyn08}.
Although popular for its simple implementation, the latter approach has two disadvantages. Firstly, a large amount of beads is needed to represent very long
or very thin fibers or chains. Secondly, the use of hard excluded volume interaction potentials necessitates small time steps to accurately integrate the equations
of motion. This makes the calculation of the dynamics of long thin rods and fibers computationally very costly.

A few off-lattice methods exist that implement non-crossing chains without resorting to explicit repulsive interactions. Examples include a Brownian dynamics
acception/rejection scheme by Ramanathan and Morse \cite{Ramanathan07} and the 'twentanglement' method of Padding and Briels \cite{Padding01,Padding02}.
Both methods, however, are based on Brownian dynamics without HIs. This means that solvent-mediated interactions between the embedded
chain segments are ignored. Rather, the segments feel a certain friction with a fictitious static background fluid, as well as random forces.

I will describe an efficient simulation algorithm for non-crossing fibers that includes hydrodynamic interactions.
The method presented here relies on the SRD method to establish HIs between fiber or chain segments.
In SRD a solvent is represented by $N_s$ ideal particles of mass $m$. After propagating the particles for a time $\delta t_c$,
the system is partitioned into cubic cells of volume $a_0^3$ (with a random grid shift to conserve Galilean invariance \cite{Ihle01}).
The velocities relative to the center of mass velocity of each
separate cell are rotated over a fixed angle around a random axis. This procedure conserves mass, momentum, and energy and yields the correct
hydrodynamic (Navier-Stokes) equations, including the effect of thermal noise \cite{Malevanets99}. The solvent particles only interact with each other
through the rotation procedure, which can be viewed as a coarse graining of particle collisions over time and
space. For this reason, the particles should not be interpreted as individual molecules but rather as a Navier-Stokes solver that naturally includes
Brownian noise. The fiber or chain segments will be coupled to this hydrodynamic solvent by also taking part in the rotation procedure. With appropriately
chosen simulation parameters \cite{Ripoll05}, such an approach leads to correct hydrodynamic behaviour of polymeric chains, as shown recently
by Winkler et al. \cite{Winkler04}. From the point of view of the latter work, this paper is an extension of the hydrodynamic method
to also include uncrossability of the chains.

This paper is organised as follows. A simple chain model is introduced in section \ref{model}. The non-crossing algorithm is described in detail
in section \ref{alg}. The choice of simulation parameters is rationalised in section \ref{parameters} and a validation and some results of the
method are given in section \ref{test}. Conclusions are given in section \ref{concl}.

\section{\label{model}Chain model}

In this work a fiber or chain is represented by a string of vertices located at positions $\RR_i$ $(i = 1,\ldots,N_v)$, with each vertex carrying a mass $M$.
The non-crossing algorithm described in the next section is generally applicable to any model in which the interactions between connected
vertices are described by potential energy terms. 
The model fiber or chain can achieve the right compressibility and bending stiffness by associating
a bonding potential energy with each bond and an angular potential energy with each bend between two successive bonds.
Specifically, the potential energy of a bond $(i,i+1)$ with length $R_{i,i+1} = | \RR_{i+1} - \RR_i |$
is given by 
\begin{equation}
\varphi^b(R_{i,i+1}) = \frac12 \frac{K}{l_0} \left( R_{i,i+1} - l_0 \right)^2.
\end{equation}
Here $K$ is the elastic modulus of the fiber or chain and $l_0$ is the equilibrium distance between each successive vertex.
Two successive bonds $(i-1,i)$ and $(i,i+1)$ with unit bondvectors $\uu_{i-1} = (\RR_i-\RR_{i-1})/R_{i-1,i}$ and
$\uu_i = (\RR_{i+1}-\RR_i)/R_{i,i+1}$ make an angle $\theta_i$ at vertex $i$, with $\cos \theta_i = \uu_{i-1} \cdot \uu_i$.
The potential energy associated with this angle is given by
\begin{equation}
\varphi^\theta(\theta_i) = k_BT \frac{l_p}{l_0} \left( 1 - \cos \theta_i \right).
\end{equation}
This particular form is chosen for the relative computational ease of calculating $\cos \theta_i$ (rather than $\theta_i$).
If $l_0 \ll l_p$ the angles will typically be small and the angular potential reduces to $\frac12 k_BT (l_p/l_0) \theta_i^2$. Such a potential ensures that the
persistence length of the fiber or chain is equal to $l_p$, as desired.
Note that more realistic (non-linear) bond and angle potentials, as well as torsional stiffness effects, may be included but are ignored for simplicity.

\section{\label{alg}Algorithm}

In order to prevent chain crossing, a rule for the detection and
treatment of bond-bond collisions needs to be devised. If hydrodynamic behaviour is to be retained, this collision rule must be compatible
with the laws of conservation of momentum and energy. The most detailed approach would be to construct an event-driven algorithm in which a list
of possible future collisions is generated and time progresses discretely from one collision instant to the next. It is possible, but rather cumbersome, to
combine such a variable timestep algorithm with the fixed timestep SRD algorithm.
However, resolving the collisions to such detail is not in the same \emph{spirit} as the SRD algorithm. In SRD one does not specify the exact locations of the collisions between
the solvent particles, but attains a rather more coarse grained view: collisions take place anywhere within the volume of a collision cell,
anytime during the collision time interval. Technically, during the collision step the solvent particles are not actually displaced, they only exchange momentum and energy.
This has proven to be sufficient for hydrodynamic behaviour of the solvent.

In this work a similar fixed timestep idea is used for the collisions between chain bonds. It is unnecessary to specify the exact locations of the collisions.
Rather, chain vertices are picked in random order and moved according to their velocities, except if this motion results in a collision with another
chain \cite{Ramanathan07}. In the latter
case momentum and energy are exchanged between the vertices surrounding the colliding bonds. By moving the chain vertices one-by-one instead of all at once, 
the detection and treatment of the collisions are greatly simplified at the cost of accuracy in the collision location.
It is necessary to use a random 
permutation for the order in which the vertices are picked, because otherwise bias may be introduced in successive collisions between
the same pair of bonds. The solvent particles of mass $m$ located at positions $\rr_j$ $(j=1,\ldots,N_s)$ are treated as usual in SRD. Both solvent particles and
chain vertices take part in the grid cell based collision step; this ensures that the chains are hydrodynamically coupled to the solvent.
The algorithm may be summarised as follows:
\begin{enumerate}
  \item Read in coordinates $(\rr, \RR)$ and velocities $(\vv, \VV)$ of the solvent particles and chain vertices.
	\item Advance solvent positions over a timestep $\delta t$
        \begin{equation}
	      \rr_j \mapsto \rr_j + \vv_j \delta t.
	      \end{equation}
	      Apply periodic boundary or wall conditions to solvent coordinates.
	\item Create a randomly permuted list of all vertices. Try moving chain vertex $\RR_i$ from this list according to
	      \begin{equation}
	      \RR_i \mapsto \RR_i^{\mathrm{trial}} = \RR_i + \VV_i \delta t.
	      \end{equation}
	      Check for crossing of the bond $(i-1,i)$ with another bond. Do the same for the bond $(i,i+1)$.
	      If a chain crossing occurs then reject this move, but exchange momentum and energy with the first collision partner.
	      If no chain crossing occurs then accept this move. Apply periodic boundary or wall conditions to chain vertex coordinates.
	      Details of crossing detection and momentum and energy exchange are given below.
	\item (May be performed less frequent:) The SRD collision step. Create a random-shifted grid and perform random collisions of solvent and chain vertices
	      within each grid cell according to
	      \begin{eqnarray}
	      \vv_j &\mapsto & \VV_{cm} + \mathbf{\mathcal{R}}(\vv_j - \VV_{cm}), \\
	      \VV_i &\mapsto & \VV_{cm} + \mathbf{\mathcal{R}}(\VV_i - \VV_{cm}).
	      \end{eqnarray}
	      Here $\VV_{cm}$ is the centre-of-mass velocity of all solvent and vertex particles in that particular cell and $\mathbf{\mathcal{R}}$ is a rotation
	      matrix which rotates velocities by a fixed angle $\alpha$ around a randomly oriented axis. Rescale velocities relative to centre-of-mass velocity if
	      thermostatting is required.
	\item Calculate vertex-vertex potential forces and possibly body forces for all particles: $\FF_i$ and $\ff_j$.
	\item Advance velocities of solvent and vertices based on forces
	      \begin{eqnarray}
	      \vv_j & \mapsto & \vv_j + \frac{\ff_j}{m} \delta t \\
	      \VV_i & \mapsto & \VV_i + \frac{\FF_i}{M} \delta t
	      \end{eqnarray}
	\item If the number of required time steps has not yet been reached, go to step 2.
	\item Save coordinates and velocities of the solvent particles and chain vertices.
\end{enumerate}

Most of the above algorithm is standard for SRD (note that in this version a leap-frog Verlet algorithm is used \cite{AllenTildesley}), except for step 3.
If the update of the positions of the chain vertices would be treated similarly to step 2, then chains would be able to cross.
More details on step 3 are given in the next subsections.

\subsection{Detecting bond crossings\label{subsec_detection}}

\begin{figure}[tb]
   \begin{center}
     \scalebox{0.6}
     {\includegraphics{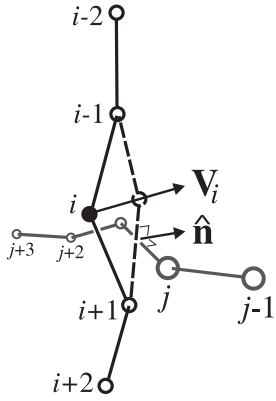}}\\
   \end{center}
   \caption{
    \label{geometry}
     When vertex $i$ is moved along its velocity vector $\VV_i$, a possible crossing of the connected bond $(i,i+1)$ with any of the
     neighbouring bonds is checked. The same applies to the connected bond $(i-1,i)$. 
     In this example a crossing between bonds $(i,i+1)$ and $(j,j+1)$ takes place. Momentum is exchanged along the direction $\nn$
     perpendicular to both these bonds at the time of impact. Note that in practice much smaller displacements of the vertices are used than
     shown here. This exaggerated view is only for reasons of clarity.
}
\end{figure}
When performing a trial move of vertex $i$ from $\RR_i$ to $\RR_i^{\mathrm{trial}} = \RR_i + \VV_i \delta t$, two bonds will move: $(i-1,i)$ and $(i,i+1)$
(see Fig.~\ref{geometry}). We assume that vertex $i$ moves linearly in time, like
\begin{equation}
\RR_i(t) = \RR_i + \VV_i t, \qquad t \in [0,\delta t].
\end{equation}
Focusing first on the bond $(i,i+1)$, an intersection of this bond with another bond $(j,j+1)$ occurs at time $t_I$ if the vectors $(\RR_{i+1}-\RR_{i}(t_I))$,
$(\RR_{i+1} - \RR_j)$ and $(\RR_{j+1}-\RR_j)$ all lie within the same plane, i.e. if
\begin{equation}
\left( \RR_{i+1} - \RR_i - \VV_i t_I \right) \cdot \left[ \left(\RR_{i+1}-\RR_j \right) \times \left( \RR_{j+1} - \RR_j \right) \right] = 0. \label{eq_tripleproduct}
\end{equation}
This may be rewritten to
\begin{equation}
t_I = \frac{\left( \RR_{i+1} - \RR_i \right) \cdot \left[ \left( \RR_{i+1}-\RR_j \right) \times \left( \RR_{j+1} - \RR_j \right) \right]}
           {\VV_i \cdot \left[ \left( \RR_{i+1}-\RR_j \right) \times \left( \RR_{j+1} - \RR_j \right) \right]}. \label{eq_tI}
\end{equation}
If $t_I \in [0,\delta t]$ a collision may have occurred. Two further checks are needed to establish whether a real collision took place between the
finite size bonds. If time is progressed to the time of intersection $t_I$, then points $\RR(s)$ on bond $(i,i+1)$ and $\RR'(s')$ on bond $(j,j+1)$ are given by
\begin{eqnarray}
\RR(s) &=& \RR_i(t_I) + s \left( \RR_{i+1} - \RR_i(t_I) \right), \ s \in [0,1] \label{eq_Rs} \\
\RR'(s') &=& \RR_j + s' \left( \RR_{j+1} - \RR_j \right), \qquad \ s' \in [0,1] \label{eq_Rsprime}
\end{eqnarray}
The point of intersection, parametrised by the pair $(s,s')$, can be found by minimising the distance $| \RR(s) - \RR'(s') |$ with respect to both parameters.
The result is
\begin{eqnarray}
s &=& \frac{be - cd}{ac - b^2}, \label{eq_s}\\
s' &=& \frac{ae - bd}{ac - b^2}, \label{eq_sprime}
\end{eqnarray}
with
\begin{eqnarray}
a &=& \left| \RR_{i+1} - \RR_i(t_I) \right|^2, \\
b &=& \left( \RR_{i+1} - \RR_i(t_I) \right) \cdot \left( \RR_{j+1} - \RR_j \right), \\
c &=& \left| \RR_{j+1} - \RR_j \right|^2, \\
d &=& \left( \RR_{i+1} - \RR_i(t_I) \right) \cdot \left( \RR_i(t_I) - \RR_j \right), \\
e &=& \left( \RR_{j+1} - \RR_j \right) \cdot \left( \RR_i(t_I) - \RR_j \right).
\end{eqnarray}
Only if both $s \in [0,1]$ and $s' \in [0,1]$ a collision has occurred between the two finite bonds, and it occurred at time $t_I$.

A similar treatment is given to the bond pair $(i-1,i)$ and $(j,j+1)$. All neighbouring bonds $(j,j+1)$ which are not directly linked
to the bonds $(i-1,i)$ or $(i,i+1)$ must be checked in this way. The use of a Verlet linked list \cite{AllenTildesley} greatly improves the 
efficiency of this procedure.

If multiple collisions occur during the time interval $[0,\delta t]$ due
to the motion of a certain vertex $i$, the first collision is chosen for the exchange of momentum and energy, as discussed in the next
subsection. 
The ratio of the number of executed collisions to the number of possible collisions is monitored during the simulations. The integration time
step should be so small that this ratio is close to one.

\subsection{Momentum and energy exchange}

Suppose that, as a consequence of the trial move of vertex $i$, a certain pair of bonds $(i,i+1)$ and $(j,j+1)$ have collided
(the case of colliding bonds $(i-1,i)$ and $(j,j+1)$ can be treated in a similar way). At the time of collision, $t_I$, an amount
$\Delta \PP$ of momentum is transferred from bond $(i,i+1)$ to bond $(j,j+1)$. 
This momentum transfer is directed along the normal to both bonds, i.e.
$\Delta \PP = \Delta P \nn$, with (see Fig.~\ref{geometry})
\begin{equation}
\nn = \frac{\left( \RR_{i+1}-\RR_i(t_I)\right) \times \left( \RR_{j+1} - \RR_j \right)}
           {\left| \left( \RR_{i+1}-\RR_i(t_I)\right) \times \left( \RR_{j+1} - \RR_j \right) \right|}.
\end{equation}

Note that in the simulation colliding bonds are not actually moved (only non-colliding bonds are). The above calculation is needed to determine the
direction in which momentum transfer is taking place.
%
Because in this model the mass is concentrated in the vertices at the extremes of the bonds, the momentum transfer must be divided between the vertices
following a lever rule. Using the fact that all vertices have the same mass $M$, the velocity change of the four vertices involved is given by:
\begin{eqnarray}
\Delta \VV_i &=& - (1-s) \frac{\Delta P}{M} \nn, \label{eq_deltaVi} \\
\Delta \VV_{i+1} &=& -s \frac{\Delta P}{M} \nn, \\
\Delta \VV_j &=& (1-s') \frac{\Delta P}{M} \nn, \\
\Delta \VV_{j+1} &=& s' \frac{\Delta P}{M} \nn. \label{eq_deltaVj1}
\end{eqnarray}
Here $s$ and $s'$ are the fractional positions along the bonds where the collision has taken place. Note that this collision automatically fulfills the
law of conservation of momentum. The \emph{amount} of momentum transfer $\Delta P$ can be found from the law of conservation of energy.
Before the momentum transfer the kinetic energy of the four involved vertices is given by
\begin{equation}
K_{\mathrm{before}} = \frac12 M \left( V_i^2 + V_{i+1}^2 + V_j^2 + V_{j+1}^2 \right),
\end{equation}
whereas after the collision it is given by
\begin{eqnarray}
K_{\mathrm{after}} &=& \frac12 M \left\{ \left| \VV_i - (1-s) \frac{\Delta P}{M} \nn \right|^2 + \left| \VV_{i+1} - s \frac{\Delta P}{M} \nn \right|^2 \right. \nonumber \\
&& \left.  + \left| \VV_j + (1-s') \frac{\Delta P}{M} \nn \right|^2 + \left| \VV_{j+1} + s' \frac{\Delta P}{M} \nn \right|^2 \right\}. \nonumber \\
\end{eqnarray}
Equating $K_{\mathrm{before}} = K_{\mathrm{after}}$ we find
\begin{equation}
\Delta P = 2M\frac{\left[ (1-s)\VV_i + s\VV_{i+1} - (1-s')\VV_j -s' \VV_{j+1}\right] \cdot \nn}{(1-s)^2+s^2+(1-s')^2+s'^2}.
\end{equation}
This is used in Eqs.~(\ref{eq_deltaVi}) - (\ref{eq_deltaVj1}) to update the vertex velocities \cite{note}.

\subsection{Extension to excluded volume fibers and chains}

The above method takes into account collisions between infinitely thin fibers or chains.
In some cases, for instance when the volume fraction is relatively
large, it is desired to take into account the excluded volume of the fibers or chains.
In this paper I will focus on the semidilute, low-volume fraction case where excluded volume is relatively unimportant (for example, the volume fraction
will be such that no spontaneous nematic ordering will occur in the equivalent experimental system). 
However, for completeness, here follows an outline of the changes that need to be made to the algorithm; a detailed account will be given in a
separate paper.

When dealing with excluded volume it is envisaged that each bond $(i,i+1)$ represents the centre line of a tube of diameter $D$. The tube stretches from $i$ to $i+1$.
Because the next bond $(i+1,i+2)$ is oriented differently, one needs to be careful at the corners. This may be done by envisaging spheres of diameter $D$ to be
placed at the vertices. When moving vertex $i$, the detection of bond crossings is more complex than the case of thin lines because the time of
collision cannot be determined independently from Eq.~(\ref{eq_tripleproduct}) anymore. Rather, a generalisation of Eq.~(\ref{eq_Rs}) is needed to indicate a point $\RR(s;t)$ on the
centre-line of bond $(i,i+1)$ at time $t$:
\begin{equation}
\RR(s;t) = \RR_i + \VV_i t + s(\RR_{i+1}-\RR_{i}-\VV_i t).
\end{equation}
Eq.~(\ref{eq_Rsprime}) is still valid to indicate a point $\RR'(s')$ on the centre-line of bond $(j,j+1)$ because this bond is not moved.
Now multiple kinds of possible collisions need to be checked: between two bonds, between a bond and a vertex, and between vertices.
The collision that has actually taken place (if any within the indicated interval) is the one with the smallest associated collision time.
These collision times are determined as follows. When checking bond $(i,i+1)$ with $(j,j+1)$, the closest distance $d_{bb}(t)$ is
determined by functionally minimising $\left| \RR(s;t) - \RR'(s')\right|$ with respect to the parameters $s$ and $s'$. The
time of impact then follows from $d_{bb}(t_I) = D$. When checking bond $(i,i+1)$ with vertex $j$, the closest distance $d_{bv}(t)$
is determined by functionally minimising $\left| \RR(s;t) - \RR_j \right|$ with respect to the parameter $s$. The time of
impact then follows from $d_{bv}(t_I) = D$. Finally, when checking vertex $i$ with vertex $j$, the closest distance $d_{vv}(t)$ is
given by $d_{vv}(t) = \left| \RR_i + \VV_i t - \RR_j \right|$. The time of impact then follows from $d_{vv}(t_I) = D$. Note that
in all these cases a grazing collision could lead to two solutions of $t_I$ within the interval $[0,\delta t]$. In that case
the smallest of the two must be considered, as that will correspond to the incoming collision.


\section{\label{parameters}Choice of parameters}

\begin{table}
\begin{center}
\caption{Units and simulation parameters for semiflexible chains in an SRD fluid.  The parameters
listed in the table all need to be independently fixed to determine a
simulation.
\label{table:units}
}
\begin{tabular}{l}
\hline \hline
 \begin{tabular}{l|l}
  \begin{tabular}[t]{l}
   \begin{tabular}{l}
      Basic Units \\
      \hline \noalign{\medskip}
   \end{tabular}  \\ \medskip

   \begin{tabular}{ll} 
      $a_0$ &= length \\  \noalign{\medskip}
      $k_B T$ & = energy \\  \noalign{\medskip}
      $m$ &= mass\\  \noalign{\medskip}
   \end{tabular} \\ \noalign{\medskip}
  \end{tabular} &
  \begin{tabular}[t]{l}
   \begin{tabular}{l}
      Derived Units \\
      \hline \noalign{\medskip}
   \end{tabular}  \\ \medskip
   \begin{tabular}{l} 
      $\displaystyle t_0 = a_0 \sqrt{\frac{m}{k_B T}}$ \hspace*{1cm} = time \\  \noalign{\medskip}
      $\displaystyle D_0 = \frac{a_0^2}{t_0} = a_0 \sqrt{\frac{k_BT}{m}}$ = diffusion constant \\ \noalign{\medskip}
      $\displaystyle \eta_0 = \frac{m}{t_0 a_0} = \frac{\sqrt{m k_B T}}{a_0^2}$ = viscosity\\  \noalign{\smallskip}
   \end{tabular} 
  \end{tabular}
 \end{tabular}
  \\
 \hline 
 \noalign{\medskip}

 \begin{tabular}{l}
  Independent fluid simulation parameters 
  \\ \hline \noalign{\medskip}
 \end{tabular}  \\ \medskip
 \begin{tabular}{ll} 
  $\gamma$ &= number of particles per cell \\  \noalign{\medskip}
  $\delta t_c$ & = SRD collision time step \\  \noalign{\medskip}
  $\alpha$ &= SRD rotation angle\\  \noalign{\medskip}
  $L_b$ &= box length\\  \noalign{\medskip}
 \end{tabular} \\

 \hline \noalign{\medskip}

 \begin{tabular}{l}
  Independent chain simulation parameters \\
  \hline \noalign{\medskip}
 \end{tabular}  \\ \medskip

 \begin{tabular}{ll} 
  $\delta t$ & = MD  time step \\  \noalign{\medskip}
  $L$ &= contour length
  \\  \noalign{\medskip}
  $l_0$ &= distance between successive vertices 
  \\  \noalign{\medskip}
  $M$ &= mass of a chain vertex
  \\  \noalign{\medskip}
  $K$ &= elastic modulus 
  \\  \noalign{\medskip}
  $l_p$ &= persistence length 
  \\  \noalign{\medskip}
  $N_c$ &= number of chains
  \\ \noalign{\medskip}
 \end{tabular} \\
 \hline \hline
\end{tabular}
\end{center}
\end{table}
Before a system of semiflexible fibers or chains in a solvent can be simulated, a number of parameters need to be chosen. A summary of these
parameters is given in Table \ref{table:units}. In this paper lengths will be in units of cell size $a_0$, energies in
units of $k_BT$, and masses in units of $m$ (this corresponds to setting $a_0 = 1$, $k_BT = 1$, and $m=1$).
Time, for example, is expressed in units of $t_0 = a_0 \sqrt{m/k_BT}$; other units can be found in Table \ref{table:units}.
The exact values of the parameters will depend of course on the particular application in mind, but there are a few general
rules which I will present here.

\subsection{Hydrodynamic coupling between the chains and the solvent}

The simulation method is supposed to capture the hydrodynamic interactions between different (parts of) chains. It is therefore important,
first, to ensure that the solvent exhibits liquidlike momentum transfer, and second to ensure a sufficiently strong coupling between
the chain vertices and the solvent.

Momentum transfer in the solvent is determined by the average number of fluid particles per cell $\gamma$, the time interval between
collisions $\delta t_c$, and the rotation angle $\alpha$. The simplicity of SRD collisions has facilitated the analytical calculation
of many transport coefficients of the solvent \cite{IhleKroll03,Kikuchi03,Pooley05,Ihle05}. These analytical expressions are particularly
useful because they enable an efficient tuning of the viscosity and other properties of the fluid, without the need for trial-and-error
simulations. The viscosity has two contributions, kinetic and collisional:
\begin{eqnarray}
\eta_{kin} &=& \frac{\gamma k_BT \delta t_c}{a_0^3} \times \nonumber \\
&& \hspace{-0.5cm} \left[ \frac{5 \gamma}{(\gamma -1 + \ee^{-\gamma})(4-2 \cos \alpha - 2 \cos 2 \alpha)} - \frac12 \right] \label{eq_etakin}\\
\eta_{col} &=& \frac{m (1-\cos \alpha )}{18 a_0 \delta t_c} ( \gamma - 1 + \ee^{-\gamma} )
\label{eq_etacol}
\end{eqnarray}
The kinetic viscosity must not be confused with the kinematic viscosity $\nu$. The latter, defined as $\nu = \eta/\rho = (\eta_{kin} + \eta_{col})/(m \gamma)$,
may be interpreted as the diffusion coefficient for momentum. In a liquid momentum diffusion is much faster than the self-diffusion
of the solvent or solute molecules (the dimensionless Schmidt number is large \cite{Padding06}).
In SRD this may be ensured by choosing the collision time interval such that the mean free path between
collisions is at least one order of magnitude smaller than the collision cell size $a_0$, i.e. $\delta t_c < 0.1 t_0$.
For a detailed treatise the reader is referred to \cite{Padding06}.
The tests described in the next section use $\delta t_c = 0.02 t_0$. 

The vertices of the chain are coupled to the solvent by participating in the collision step. Ripoll et al. \cite{Ripoll05} have shown that
an optimal hydrodynamic coupling is achieved when the mass of the vertex is about equal to the total mass of the solvent
particles in a cell and, as above, when the collision interval is chosen sufficiently small. The tests described in the next section use 
$\gamma = 5$ and $M = 5m$. Under these conditions the selfdiffusion $D_M$ of the vertex is for a large
part determined by hydrodynamic correlations in the solvent. The effective hydrodynamic radius, defined as $a_h = k_BT/(6 \pi \eta D_M)$, is
approximately $0.3 a_0$. The hydrodynamic interactions between different (segments of)
fibers will then be correctly reproduced if the equilibrium distance $l_0$ between connected vertices is about twice the hydrodynamic radius.
Similar to the work of Winkler et al. we choose $l_0 = 0.5 a_0$, which for flexible polymer chains was shown to yield the expected
Zimm dynamics \cite{Winkler04}.

The value of the rotation angle $\alpha$ also determines the amount of hydrodynamic coupling \cite{Ripoll05}.
Obviously, the coupling will be less for smaller rotation angles; in the limit $\alpha = 0$ no momentum will be transfered between chain and solvent.
Generally, in the range $\pi/2 \leq \alpha < \pi$ the exact value of $\alpha$ is much less important for the coupling than the value of the
collision interval (note that extremes near $\alpha = \pi$ should be avoided).
Since rotations around an angle of $\alpha = \pi/2$ can be implemented particularly efficiently, this value was chosen in all work described here.

The SRD method has proven to be very robust when it comes to predicting hydrodynamic behaviour of embedded objects, in both equilibrium and
nonequilibrium situations \cite{Kikuchi03,Winkler04,Padding04,Padding05,Ripoll05,Padding06,Goetze07,Watari07,Padding08}.
The precise speed of the dynamics depends on the choice of the above parameters, just as in a real experiment choosing glycerine instead of water
will slow down the dynamics of embedded objects. Some choices will be computationally more efficient than others but as long as the appropriate
limits mentioned above ($\delta t_c < 0.1 t_0$, $\pi/2 \leq \alpha < \pi$, $M \approx \gamma m$, and $l_0 \approx a_0$) are respected,
the physical hydrodynamic behaviour of the system will be correctly simulated.

\subsection{Dimensionless numbers}

\begin{table}[tb]
	\centering
	\caption{Dimensionless numbers relevant to simulation of the statics and dynamics of a fiber suspension and
	the specific values used in the test simulations. When flow is applied (not in this work), one should be 
	particularly mindful of the Mach and Reynolds numbers, which are reported here for the case of shear flow with shear
	rate $\dot{\gamma}$. Recommended upper limits for Stokes flow are given. The Peclet
	number may be smaller or larger than 1, depending on experimental conditions.}
	\label{table_parameters}
		\begin{tabular}{llr}
		\hline \hline
		property & definition \ & value \\
		\hline
		dimensionless persistence length & $l_p^* = l_p/L$ & 1 \\
		hydrodynamic aspect ratio & $p = L/(2a_h)$ & 64 \\
		dimensionless mesh size & $\xi^* = \xi/L$ & 0.055 - 0.32 \\
		Compressibility effects & $\mathrm{Ma} = \dot{\gamma}l_p/c_s$ & $< 0.1$ \\
	  Inertial vs. viscous forces & $\mathrm{Re} = \dot{\gamma}l_p^2/\nu$ & $< 0.1$ \\
	  Convective vs. Brownian motion & $\mathrm{Pe} = \dot{\gamma} \tau_r$ & \\
	  \hline \hline
		\end{tabular}
\end{table}
Tuning of the model to experimental conditions is greatly facilitated by the use of dimensionless numbers. The dimensionless numbers which are
relevant to  a fiber suspension are summarised in Table \ref{table_parameters}. The ratio of persistence length $l_p$ to fiber contour length $L$,
\begin{equation}
l_p^* = \frac{l_p}{L},
\end{equation} 
determines whether the fibers are flexible ($l_p^* \ll 1$), semiflexible ($l_p^* \approx 1$) or stiff ($l_p \gg 1$).
The aspect ratio of the fiber 
\begin{equation}
p = \frac{L}{2a_h},
\end{equation}
where $a_h$ is the (hydrodynamic) radius, is important for the hydrodynamic behaviour of the fiber.
For example, the rotational and translational diffusion coefficient of a stiff rod depend strongly on $p$, even in dilute solutions \cite{Tirado84,DoiEdwards},
and the critical concentration for nematic ordering due to excluded volume depends on the ratio between
persistence length and diameter, i.e. on $p l_p^*$ \cite{DoiEdwards}.

A network of fibers is further characterised by its mesh size:
\begin{equation}
\xi \equiv \sqrt{\frac{3}{cL}}.
\end{equation}
Here $c$ is the number density of fibers. The mesh size can be interpreted as an average distance between network segments, where
the numerator 3 is a mere definition. An important dimensionless number is the ratio of mesh size to contour length $\xi^* = \xi/L$.
Together with the dimensionless persistence length, it determines the amount of confinement that a fiber feels due to entanglements with
its neighbours \cite{Morse01,Ramanathan07,Hinsch07}. For example, Hinsch et al. \cite{Hinsch07} derive an effective tube diameter $L_{\perp}$ given by
\begin{equation}
\frac{L_{\perp}}{L} = 0.31 \frac{\left(\xi^*\right)^{6/5}}{\left(l_p^*\right)^{1/5}} + 0.56 (\xi^*)^2, \label{eq_tubed}
\end{equation}
and a deflection length $L_d$ (average distance between successive collisions of the fiber
with its tube) $L_d/L = 0.64 (\xi^*)^{4/5} (l_p^*)^{1/5} + 0.39 (\xi^*)^{8/5} (l_p^*)^{2/5}$.
These expressions confirm the importance of the dimensionless numbers $l_p^*$ and $\xi^*$. Note that Eq.~(\ref{eq_tubed})
confirms the established scaling law $L_{\perp} \propto \xi^{6/5} l_p^{-1/5}$, valid for long enough chains \cite{Odijk83,Semenov86,Morse98}.

When flow is applied (this will be presented in a forthcoming article), a few more dimensionless numbers need to be taken into account to correctly
characterise the relative strength of competing physical processes \cite{Guyon01}.
Firstly, the Mach number measures the ratio
\begin{equation}
\mathrm{Ma} = \frac{v_{flow}}{c_s},
\end{equation}
between $v_{flow}$, the (relative) flow speed of the solvent, and $c_s = \sqrt{(5/3)(k_BT/m)}$, the speed of sound. The Mach number measures compressibility
effects \cite{Guyon01} since the sound speed is related to the compressibility of a liquid. 
It may sound obvious that Ma needs to remain small ($\ll 1$) for physical fiber suspensions, but particle-based coarse-graining schemes drastically
increase the Mach number. The fluid particle mass $m$ is typically much greater than the mass of a molecule of the underlying fluid, resulting in
a lower speed of sound. In other words, particle based coarse-grained systems are typically much more compressible than the solvents they model.
In practice, in order to avoid compressibility effects in the dynamics of the system, the Mach number must remain lower than about 0.1 \cite{Padding06}.

The Reynolds number is one of the most important dimensionless numbers characterising hydrodynamic flows. Mathematically, it measures the relative
importance of the non-linear terms in the Navier-Stokes equation \cite{Guyon01}. Physically, it determines the relative importance of inertial over
viscous forces and can be expressed as
\begin{equation}
\mathrm{Re} = \frac{v_{flow}R}{\nu},
\end{equation}
where $R$ is a length scale relevant to the problem. For a fiber suspension this could be the persistence length, i.e. $R \approx l_p$.
For micrometer sized objects, the Re is usually very small ($\mathrm{Re} \ll 1$). The Reynolds number can be kept small by ensuring that the flow
velocities do not exceed some maximum (this should be monitored during the simulation) and by choosing a relatively high kinematic viscosity. Again, the latter
may be done by choosing a small collision interval $\delta t_c$.

The definitions of the Mach and Reynolds numbers above depend on the chosen relevant length scale as well as the characteristic flow velocity.
In Table \ref{table_parameters} we report Ma and Re for shear flow with shear rate $\dot{\gamma}$,
where the relevant length scale of a semiflexible fiber is the persistence length,
and the characteristic flow velocity is the maximal velocity difference over this length scale.

Lastly, it is important that the relative importance of convective transport to diffusive transport is comparable between experiment and
simulation. This is expressed by the Peclet number
\begin{equation}
\mathrm{Pe} = \frac{v_{flow} R}{D},
\end{equation}
where $D$ is the self-diffusion coefficient of the fiber. Alternatively, under shear flow the Peclet number
can be defined as the product of applied shear rate and the (Brownian) rotational relaxation time $\tau_r$ of the fiber or chain.
In this respect it should be noted that the absence of excluded volume interactions facilitates the simulation of very long and thin fibers,
with very large characteristic times. 
For example, the characteristic times associated with rotational, perpendicular and parallel motion of a stiff rod scale like
$L^3/[\ln p + f(p)]$, with $f(p)$ weak functions of the aspect ratio $p$ \cite{Tirado84}.
In the large $p$ limit, the friction perpendicular to the rod is twice that in the parallel direction.
This large $p$ limit is reached (within a few percent accuracy) for $p$ in the order of 30 \cite{Dhont}. Therefore, a connection between
time in a simulation of rods with $p = 30$ and time in an experiment with much longer rods can be made by identifying the rotational
relaxation time of the simulated rods with the rotational relaxation time in the experiment.

\subsection{Galilean invariance}

Although momentum is conserved locally in all solvent and bond collisions, the method presented here is not
strictly Galilean invariant.
Remember that a bond which exchanges momentum with its collision partner is not actually displaced.
In this step reference is made to an absolute reference frame: the centre-of-mass of the collision partners should have been displaced over
a distance $\VV_{cm} \delta t$, where $\VV_{cm}$ is the centre-of-mass velocity of the collision partners. The influence of neglecting
this centre-of-mass motion in one time step can be made arbitrarily small by choosing a sufficiently small molecular dynamics step $\delta t$.
When the \emph{number} of time steps in which a particular bond collides is much smaller than the number of time steps in which the bond is
moving according to its given velocity, correct dynamics is recovered.

As it turns out, the above condition is not limiting the efficiency
of the method, for three reasons. Firstly, the ratio of bond length $l_0$ to mesh size $\xi$ is usually small, making collisions relatively
rare for each particular bond. Secondly, the molecular dynamics time step $\delta t$ already needs to be chosen relatively small to resolve
the dynamics of the relative stiff springs needed to represent real fibrillar materials, such as actin. Thirdly, the Mach number limit
introduced above already limits the allowed flow velocities, and hence the magnitude of $\VV_{cm}$.
Typically these limits imply $\left| \VV_{cm} \right| < 0.1 a_0/t_0$ and $\delta t < 0.01 t_0$,
i.e. the error in the update of the centre-of-mass position of two colliding bonds is less than $0.001 a_0$.
This is much smaller than any of the other typical length scales ($L,l_p,\xi,l_0$) of the problem.
In the next section a test will show that the method is indeed effectively Galilean invariant for all tested flow velocities 
in the range $0 \leq v_{flow} \leq 0.54 a_0/t_0$.

\subsection{Computational efficiency}

The ability to update the positions and check for collisions one vertex at a time makes the method efficient. Also, the use of
MPCD to model the solvent makes the inclusion of hydrodynamic interactions relatively cheap.
The precise speed of the simulation depends on the system size and chain density, where the computation rate scales approximately inversely linear 
with system volume and $cL^3$.
In its current implementation a system containing about $1.6 \times 10^5$ solvent particles and 100 semiflexible fibers with an aspect
ratio of $p = 64$ (i.e. represented by 64 vertices each) at a density of $cL^3 = 100$ is integrated at a rate of 30 time steps ($\delta t$) per second
on a modern single core processor.
For this system one (dilute limit) rotational relaxation time $\tau_r \approx 8.3 \times 10^4\, t_0$ is reached in 75 hours.
Of course $\tau_r$ itself depends strongly on the length of the fiber.
In the above example, when each 64 vertex fiber is cut into two shorter fibers of 32 vertices, using the same mesh size, the time to
reach $\tau_r$ decreases to 11 hours of computation.

\section{\label{test}Validation and results}

\subsection{Dilute chains and fibers}

The dynamics of a flexible chain or semiflexible fiber in dilute solution is strongly affected by HIs. To test whether
the SRD method indeed captures hydrodynamic interactions correctly, I will first focus on the qualitative and quantitative
behaviour of the self-diffusion coefficients of single chains or fibers \cite{Winkler04}.
In all cases the solvent is represented by an average of $\gamma = 5$ particles per cell.
The collision interval is set to $\delta t_c = 0.02 t_0$ and the collision angle to $\alpha = \pi/2$.

\begin{figure}[tb]
   \begin{center}
     \scalebox{0.45}
     {\includegraphics{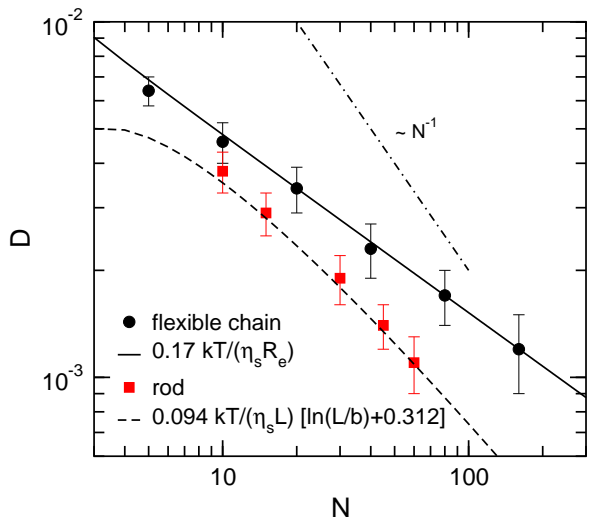}}\\
   \end{center}
   \caption{
    \label{Dscaling}
     Simulation results for the center of mass diffusion coefficient (in units $a_0^2/t_0$) for dilute flexible chains (black circles)
     and rigid-like rods ($L_p = 2L$; red squares) for various contour lengths. The solid and dashed lines correspond to theoretical
     predictions Eqs.~(\ref{eq_Dchain}) and (\ref{eq_Drod}). The dot-dashed line indicates the scaling expected for chains or rods
     without hydrodynamic interactions.
}
\end{figure}
Flexible chains are represented by $N = 5, 10, 20, 40, 80$ or 160 vertices of mass $M = 5m$ at an equilibrium distance $l_0 = 0$ and a bond strength
$k = 3\,k_BT/a_0^2$, corresponding to an entropic spring with root-mean-square bond length $l = 1 a_0$ \cite{DoiEdwards} (all angular interactions
have been disabled).
The size of the cubic periodic simulation box is varied linearly with the root-mean-square end-to-end distance $R_e = l\sqrt{N}$ to avoid artifacts
due to finite system sizes. Explicitly, $L_b = 25 a_0$ is chosen for $N = 10$.
For flexible chains, hydrodynamic Zimm theory predicts a self-diffusion coefficient given by \cite{DoiEdwards}
\begin{equation}
D = 0.196 \frac{k_BT}{\eta R_e} \qquad \mbox{(flexible chain)}\label{eq_Dchain}
\end{equation}
Fig.~\ref{Dscaling} presents the diffusion coefficients of the centres of mass of flexible chains of various length (black circles).
Qualitatively, a scaling $D \propto R_e^{-1} \propto N^{-1/2}$ can be observed.
Quantitatively, using the analytically known viscosity from Eqs.~(\ref{eq_etakin}) and (\ref{eq_etacol}), good agreement is found if the prefactor 0.196 in
Eq.~(\ref{eq_Dchain}) is replaced by 0.17 (solid line). A slightly lower self-diffusion is in agreement with the fact that periodic images of the chain
interact with each other via the periodic boundaries \cite{Padding06}.

For rigid rods, the self-diffusion coefficient is given by \cite{Tirado84}
\begin{equation}
D = \frac{k_BT}{3 \pi \eta L} \left[ \ln\left( \frac{L}{b} \right) + 0.312 \right] \qquad \mbox{(rigid rod)}\label{eq_Drod}
\end{equation}
(up to order $b/L$), where $b$ is the hydrodynamic diameter of the rod. To verify this relation, single rod-like fibers are
represented by $N = 10, 15, 30, 45$ or 60 vertices of mass $M = 5m$ at an equilibrium distance
$l_0 = 0.5 a_0$ and bond springs with strength $k = K / l_0 = 100\, k_BT/a_0^2$. In order to minimise effects of flexibility, the persistence length
is chosen equal to twice the contour length, $l_p = 2L$. The relatively stiff bonds and angles require a molecular dynamics integration step of
$\delta t = 0.01 t_0$. To avoid artifacts in the determination of the fiber length dependence due to finite system size effects, the size of the
cubic periodic simulation box is increased linearly with the length of the fiber, where $L_b = 18 a_0$ is chosen for $N=10$.
Fig.~\ref{Dscaling} presents the diffusion coefficients of the centres of mass of rod-like fibers of various length (red squares), 
together with the theoretical curve Eq.~\ref{eq_Drod} (dashed line). Similar to the work described in \cite{Winkler04} the diameter
$b$ and the prefactor are obtained by a least squares fit, yielding $b = 0.6 a_0$ and a prefactor 0.094. The diameter is in good agreement with
the effective hydrodynamic radius estimated for our vertices. The prefactor is slightly smaller than the theoretical prediction $1/(3 \pi) = 0.106$,
which can again be attributed to the slowing effect of periodic images.

Note that in the absence of hydrodynamic interactions each vertex would act as an independent source of friction, leading to a 
centre-of-mass diffusion coefficient which scales like $N^{-1}$ (dot-dashed line) for both flexible chains and rigid rods. 
From these tests it may be concluded that the SRD method correctly captures the hydrodynamic interactions for flexible chains and semiflexible fibers.

\subsection{Semidilute fibers}

In the following tests I will focus on the dynamics of suspensions of many semiflexible fibers, each similar to the rod-like fibers studied above, but
now represented by 64 vertices and a persistence length equal to the contour length, i.e. $l_p = L = 32 a_0$.
All simulations were performed in a periodic cubic box with sides $L_b = 32 a_0$. The number density $c$ of fibers was varied between the values
$cL^3 = 30$, 100, 300 and 1000, corresponding to mesh sizes $\xi = 10.2 a_0$, $5.44 a_0$, $3.20 a_0$ and $1.76 a_0$, respectively.
Higher values of the fiber density are not relevant because excluded volume effects can then no longer be neglected \cite{DoiEdwards}.

\subsubsection{Validation of Galilean invariance}

To test the effective Galilean invariance 
of the non-crossing constraint
explicitly, a periodic system of semiflexible fibers at the highest density of $cL^3 = 1000$ 
was subjected to a homogeneous flow in the $x$-direction with velocities ranging from zero to a relatively high $v_{flow} = 0.54\, a_0/t_0$.
During a run of $10^5$ integration steps, several properties were monitored and compared to a system at rest ($v_{flow} = 0$).

The energy and the centre-of-mass velocity of the system was observed to remain exactly constant. This confirms that energy and momentum are conserved during the
fiber collisions also in the presence of background flow.

\begin{figure}[tb]
   \begin{center}
     \scalebox{0.45}
     {\includegraphics{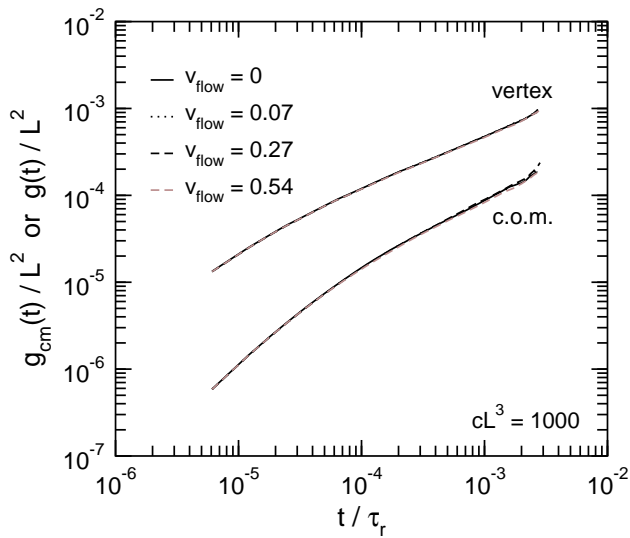}}\\
   \end{center}
   \caption{
    \label{msd_N1000}
     Mean square displacement of the vertices $g(t)$ and the fiber centres-of-mass $g_{cm}(t)$, normalised by the square of the contour length $L$
     of the fiber, measured relative to a background flowing with velocity $v_{flow}$ (in units $a_0/t_0$) in the $x$-direction,
     in a system with $l_p = L$ and $cL^3 = 1000$. Time is normalised by
     the rotation time $\tau_r$ of a fiber in the dilute limit. The results are indistinguishable for all flow velocities up to $0.54\, a_0/t_0$,
     signifying effective Galilean invariance of the method.
}
\end{figure}
The vertex mean square displacement
\begin{equation}
g(t) = \left\langle \left( \RR_i(t) - \RR_i(0) \right)^2 \right\rangle
\end{equation}
averaged over all vertices $i$, as well as the mean square displacement
\begin{equation}
g_{cm}(t) = \left\langle \left( \RR_{cm}(t) - \RR_{cm}(0) \right)^2 \right\rangle
\end{equation}
of the fiber centres-of-mass (both relative to the background flow) were determined and observed to be nearly indistinguishable,
as shown in Fig.~\ref{msd_N1000}. This conclusively shows that, for the chosen parameters, the method is effectively Galilean invariant
for all relevant flow velocities.

\subsubsection{Influence of hydrodynamic interactions and uncrossability of fibers}

Hydrodynamic interactions may easily be turned off by selecting random pairs of fluid particles after the collision step and exchanging their velocities.
In this manner energy and momentum are still conserved globally, but no longer locally. The non-crossing constraint can be turned off by simply skipping the
bond collision check. 

\begin{figure}[tb]
   \begin{center}
     \scalebox{0.45}
     {\includegraphics{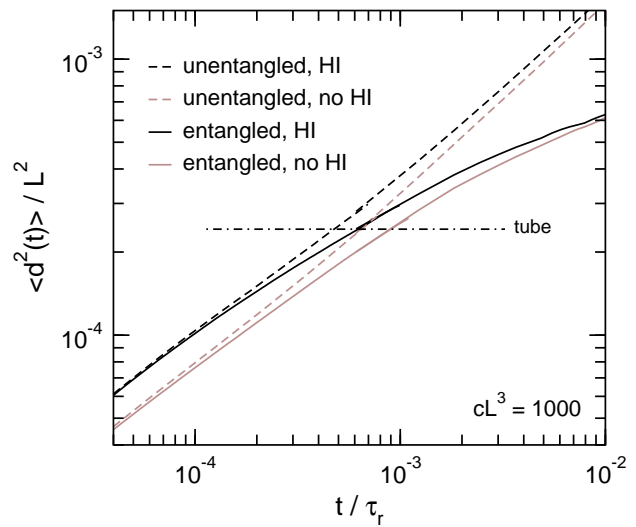}}\\
   \end{center}
   \caption{
    \label{vard_N1000}
     Mean square displacement $\left\langle d^2(t)\right\rangle$ of the minimum distance between the middle vertex at time $t+\tau$
     and the fiber at time $\tau$, without the non-crossing constraint (unentangled, dashed lines) and with the non-crossing
     constraint (entangled, solid lines), and with hydrodynamic interactions (HI, black colored lines) and without 
     hydrodynamic interactions (no HI, grey lines). The dash-dotted line shows the tube diameter predicted by Eq.~(\ref{eq_tubed}).
     For all data shown $l_p = L$ and $cL^3 = 1000$.
}
\end{figure}
Figure \ref{vard_N1000} shows the effect of hydrodynamic interactions and uncrossability of fibers on a quantity $\left\langle d^2(t) \right\rangle$, where $d(t)$
is defined as
\begin{equation}
d(t) = \min_{j} \left| \RR_m(t+\tau) - \RR_j(\tau) \right|,
\end{equation}
where $m = N_v/2$ is the middle vertex of a fiber and $j$ runs over all vertices ${1,\ldots,N_v}$ of that fiber. In other words, $d$ is
the closest distance between the position $\RR_m$ of the middle vertex at time $t+\tau$ and any of the vertices of the fiber at an
earlier time $t$. In a tightly entangled solution, the magnitude of the plateau in this quantity is a measure of the width of the tube to which
the fiber is confined \cite{Ramanathan07,Ramanathan07b}.
The time axis is normalised by the rotation time $\tau_r$ of a fiber in the dilute limit, measured from the end-to-end
vector decorrelation of a single fiber in a box of the same dimensions and with or without HIs, respectively.
Two observations can be made.

First, the results without HIs (grey lines) are systematically below the results with HIs (black lines). The relative
difference is larger at shorter correlation times than at longer correlation times, leading to small differences in scaling of 
$\left\langle d^2(t)\right\rangle$ with time $t$. It may be concluded that, apart from such small differences, the overall behaviour with or without HIs
is quite similar for semiflexible fibers of length $L = l_p$. This result is in agreement with findings for completely rigid rods ($l_p \gg L$)
where it was found that the effects of HIs are secondary relative to the steric interactions \cite{Pryamitsyn08}. 

Second, the results using the non-crossing constraint (solid lines) are equal to the results without this constraint (dashed lines) at short times,
whereas they deviate significantly at larger times. The transition between these two regimes may be interpreted in the tube model \cite{DoiEdwards} as the
moment when the fibers start to collide with their effective tube walls. Fig.~\ref{vard_N1000} shows the prediction $2L_{\perp}^2$ of
Eq.~(\ref{eq_tubed}) (horizontal dash-dotted line labeled `tube'), where the factor of 2 arises because in the theory of Ref. \cite{Hinsch07} 
$L_{\perp}$ is defined as the mean square transverse displacement of one Cartesian component only. The agreement between the observed and predicted
tube diameter is good.

\begin{figure}[tb]
   \begin{center}
     \scalebox{0.45}
     {\includegraphics{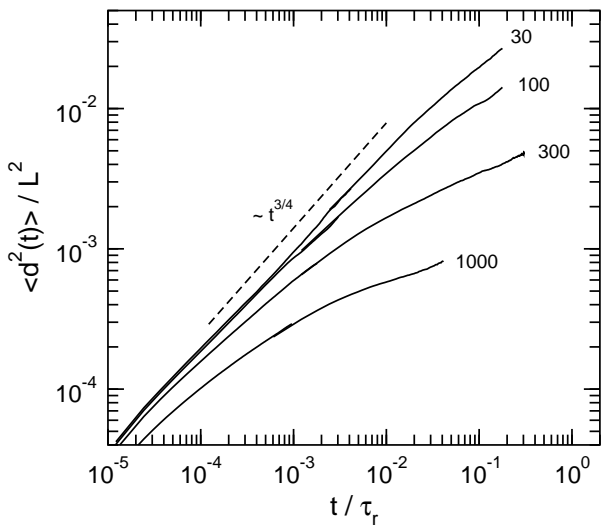}}\\
   \end{center}
   \caption{
    \label{vard_nc_h}
     Mean square displacement $\left\langle d^2(t)\right\rangle$ of the minimum distance between the middle vertex at time $t+\tau$
     and the fiber at time $\tau$ with the non-crossing constraint and hydrodynamic interactions for concentrations $cL^3 = 30$, 100, 300 and
     1000. The dashed line indicates the expected $t^{3/4}$ behaviour for a single semiflexible fiber. For all data shown $l_p = L$.
}
\end{figure}
Focusing now on the most realistic case, with HIs and non-crossing fibers, the influence of network density is shown in Fig.~\ref{vard_nc_h}. For
the lowest density shown, $cL^3 = 30$, the fibers behave almost as dilute single fibers. In this limit, the growth of
$\left\langle d^2(t)\right\rangle$ with time $t$ at early times is limited by the finite transversal fluctuations of a wormlike chain, leading to
an expected $t^{3/4}$ scaling \cite{Granek97}. This is indeed observed in the simulations as well (dashed line). With increasing network density (and hence
decreasing mesh size) the displacement of the fibers become hindered by the presence of other fibers at smaller and smaller length scales.
A more in-depth analysis will be presented in a forthcoming paper.

\section{\label{concl}Conclusions}

I have introduced a method to simulate the dynamics of Brownian fiber suspensions, where hydrodynamic interactions are mediated by a mesoscopic solvent
and collisions between fibers are treated such that momentum and energy are conserved locally. The method is made efficient by moving one fiber segment at
a time instead of all segments at once. A similar idea was used in the work of Ramanathan and Morse \cite{Ramanathan07} in the context of non-hydrodynamic
Brownian dynamics, whereas in this work hydrodynamics are conserved. The effective Galilean invariance of the current method was explicitly checked.

It was found that for 
semidilute
semiflexible fibers with $L = l_p$ the effects of hydrodynamic interactions are small compared to the effects of
uncrossability of the fibers. Because a similar observation has already been made for completely rigid rods \cite{Pryamitsyn08}, it may be
concluded that HIs are relatively unimportant for all 
semidilute suspensions of
fibers for which $L \le l_p$. This is also the reason why the observed displacements of fibers
in a hydrodynamic solvent are globally similar to those obtained in non-hydrodynamic simulations \cite{Ramanathan07,Ramanathan07b}, although
differences are observed upon closer inspection. 
At constant chain concentration,
these differences will become increasingly more important for longer chains ($L > l_p$) \cite{Winkler04},
or in situations where fibers are subjected to flow. The purpose of this paper was to introduce and validate the method;
in a forthcoming paper I will focus on non-equilibrium situations. For example, the effect of viscous drag and hydrodynamic interactions will be studied
in microrheology experiments where the response of an actively driven probe bead in a fiber suspension is measured.

\begin{acknowledgments}
This work was financed by the Netherlands Organisation for Scientific Research (NWO) through a VENI grant. Wim Briels is acknowledged for stimulating discussions.
\end{acknowledgments}

\end{document}